\documentclass[showkeys,aps,10pt,twocolumn,showpacs,preprintnumbers,amsmath,amssymb,prd,letterpaper,floatfix,nofootinbib,superscriptaddress,]{revtex4-1}
\usepackage{graphics}
\usepackage{graphicx}
\usepackage{epsfig}
\usepackage[usenames]{xcolor}
\usepackage{longtable}
\PassOptionsToPackage{hyphens}{url}\usepackage{hyperref}
\usepackage{soul}
\usepackage{breakurl}
\usepackage{latexsym}
\usepackage{amsmath}
\usepackage{amsthm}
\usepackage{amsfonts}
\usepackage{amssymb}
\usepackage{dsfont}
\usepackage{bm} 

\newcommand{\be}{\begin{equation}}
\newcommand{\ee}{\end{equation}}
\newcommand{\bel}{\begin{align}}
\newcommand{\eel}{\end{align}}
\newcommand{\bea}{\begin{eqnarray}}
\newcommand{\eea}{\end{eqnarray}}

\newcommand{\I}{\mathrm{i}}  
\newcommand{\E}{\mathrm{e}}  

\newcommand{\RE}{\mathbb{R}}
\newcommand{\ot}{\ensuremath{\frac{1}{2}}}
\newcommand{\tr}{\mathrm{tr}}

\newcommand{\psibar}{\overline\psi}

\newcommand{\psidag}{{\psi^\dagger}}

\newcommand{\eq}[1]{(\ref{#1})}

\newcommand{\ubar}{\overline{u}}
\newcommand{\dbar}{\overline{d}}
\newcommand{\psiR}{R}
\newcommand{\psiL}{L}
\begin{document}

\title{ \texorpdfstring{Low lying eigenmodes and meson propagator symmetries }{}}

\author{C.~B.~Lang}
\email{christian.lang@uni-graz.at}
\affiliation{Institute of Physics,  University of Graz, A--8010 Graz, Austria}

 \date{\today}

\begin{abstract} 
In situations where the low lying eigenmodes of the Dirac operator are suppressed one observed degeneracies of some meson masses. Based on these results a hidden symmetry was conjectured, which is not a symmetry of the Lagrangian but emerges in the quantization process. We show here how the difference between classes of meson propagators is governed by the low modes and shrinks when they disappear. 
 \end{abstract}
 

\maketitle

\section{Motivation}

Recently it was found that in certain situations a symmetry emerges that relates vector and scalar meson propagators but that is no symmetry of the action. That symmetry was observed in lattice QCD when low lying eigenmodes of the Dirac operator are suppressed either artificially by removing the eigenmodes from the quenched quark propagators \cite{Lang:2011vw,Glozman:2012fj,Denissenya:2014poa,Denissenya:2014ywa,Denissenya:2015mqa} or naturally in the high temperature phase \cite{Tomiya:2016jwr,Rohrhofer:2017grg} either due to a gap\footnote{The existence of a gap above the finite temperature transition in lattice QCD is disputed \cite{Dick:2015twa,Tomiya:2014mma,Ohno:2012br,Kovacs:2008sc}.} or because another rapid decrease towards zero eigenvalues. The symmetry group was called CS (chiral-spin) and has been suggested \cite{Glozman:2015qva,Glozman:2014mka}  to be $SU(4)\supset SU(2)_L\times SU(2)_R\times U(1)_A$ mixing the $u$- and $d$-quarks of a given chirality and also the left- and right-handed components.
 
Here we  provide results to elucidate the observed symmetry. We find explicitly in an analytic calculation that some propagator identities emerge if the low lying eigenmodes of the Dirac operator are suppressed. We show that in that case propagators of different mesons become degenerate giving rise to the CS symmetry in part. For the other part of the symmetry further conditions have to be met.

Consider the eigenvalues of the hermitian  Dirac operator. It is well known that the difference of the susceptibilities of, e.g., the propagator of the isovector scalar meson operator and of the isovector pseudoscalar operator are weighted by an eigenvalues density factor (on top of the generic eigenvalue density), that approaches a delta function in the zero mass limit. The approach is similar to the derivation of the Banks-Casher relation for the quark condensate \cite{Banks:1979yr}. The difference between the scalar and pseudoscalar propagators and susceptibilities has been intensely studied earlier \cite{Cohen:1996ng,Aoki:2012yj,Fukaya:2017wfq}. We show here that this property applies to  a large set of (scalar, pseudoscalar, vector and axial vector) meson propagator pairs and discuss the conditions for the CS symmetry.
\section{Notation}

\subsection{Dirac operator}

We work in Euclidean space-time continuum and will briefly remind on the notation. The tool will be the spectral representation of the  Dirac quark and the meson propagator. As there are sums over all (an infinite number of) eigenmodes we need some regularization (e.g., a finite volume lattice) and for this we rely on Fujikawa's approach \cite{Fujikawa:1979ay,Fujikawa:1980eg,Fujikawa:2004}, which we assume implicitly but will omit the actual derivation.

We choose hermitian $\gamma$-matrices $\gamma_\mu^\dagger=\gamma_\mu$ and $[\gamma_\mu,\gamma_\nu]_+=2\delta_{\mu\nu}$. The fermions are Grassmann fields and the Dirac action
\be
\int d^4x \psibar ( \I \gamma_\mu D_\mu+\I m) \psi
\ee
is real.
The massless Euclidean Dirac operator $D\equiv \I\gamma_\mu D_\mu$ is hermitian with the eigensystem
\be\label{eq:O1basis}
D\psi^{(n)}_{x,a}=\eta_n  \psi^{(n)}_{x,a}
\ee
The dimension of the eigenvectors is $n_D n_c n_f$ (Dirac, color, flavor) at each point $x\in\RE^4$ and there are $n_D n_c n_f$ eigenvectors as functions of $x$. Only the Dirac index $a$ is kept explicitly, the color- and flavor indices are implicit. The non-zero eigenvalues come in pairs as can be seen by multiplying with $\gamma_5$:
\bea
\gamma_5 D \psi^{(n)}&=&\eta_n \gamma_5 \psi^{(n)}\\
\quad\to\quad
D  (\gamma_5 \psi^{(n)})&=&-\eta_n  (\gamma_5 \psi^{(n)})\;.\nonumber
\eea
We use the notation $\eta_n\equiv \eta_n$ with $\eta_{-n}=-\eta_n $ and $\psi^{(-n)}=\gamma_5\psi^{(n)}$. 
The eigenvalues are real and the eigenvectors form an orthonormal basis
\be
\sum_a \int d^4x\; \psi^{(n)\dagger}_{xa} \psi^{(k)}_{xa}=\delta_{nk}\;.
\ee
We formally regularize by point-splitting such that the Dirac operator becomes a matrix,
\be
D_{xa|yb}\psi_{yb}^{(n)}=\eta_n\psi_{xa}^{(n)}\;.
\ee
The indices for color and flavor are implicit. For simplicity we assume mass degenerate fermions. The Dirac operator has the spectral representation
\be
D_{xa|yb}+\I \,m \delta_{xa|yb}=\sum_n (\eta_n+\I \,m)\psi^{(n)}_{xa}\psi^{(n)\dagger}_{yb}\;.
\ee
The fermion propagator then has the representation
\be
(D+\I\, m)^{-1}_{xa|yb}=\sum_n \frac{1}{\eta_n+\I \,m}\psi^{(n)}_{xa}\psi^{(n)\dagger}_{yb}\;.
\ee
We assume that there are no exact zero modes with $\gamma_5 \psi=\pm \psi$ either because we are in the
topological sector zero or because they have been removed. Anyhow they are suppressed in the thermodynamic limit.

\subsection{Chiral symmetry}\label{sec_chiral_symmetry}

We define chiral symmetry as {\em the invariance of the massless Dirac operator.} The transformation is
\bea\label{chiralTrans}
&\psi(x)'= \E^{\I\alpha \gamma_5\mathbf{1}_f}\psi(x)\;,\quad  &\psibar(x)'=\psibar(x) \E^{\I\alpha \gamma_5\mathbf{1}_f}\;,\\
&\psi(x)'= \E^{\I\alpha \gamma_5\tau_i}\psi(x)\;,\quad  &\psibar(x)'=\psibar(x) \E^{\I\alpha \gamma_5\tau_i}\;,
\eea
where $\tau_i$ are the generators of $SU(2)_{flavor}$ and $\mathbf{1}_f$ the unit matrix in flavor space. The kinetic term of the action is invariant, e.g.,
\be
\psibar\gamma_\mu\psi \to
\psibar \E^{\I\alpha\gamma_5}\gamma_\mu \E^{\I\alpha\gamma_5}\psi=
\psibar\gamma_\mu\psi
\ee
The chiral transformation commutes with the Euclidean Lorentz transformations $O(4)$.

For the discussion it will be useful  to split the four-Dirac-components of the eigenvectors into a pair ,
\be
\psi^{(n)}=\psi_{R}^{(n)}+\psi_{L}^{(n)}
\ee
with
\be
\psi_{R}^{(n)}=\ot(\mathbf{1}+\gamma_5)\psi\;,\;\;\psi_{L}^{(n)}=\ot(\mathbf{1}-\gamma_5)\psi\;.
\ee
We choose a chiral basis for the Dirac matrices with $\gamma_5=\mathrm{diag}(1,1,-1,-1)$
such that
\be\label{LRbasis}
\psi_{R}^{(n)}=\left(\begin{matrix}\psiR^{(n)}\cr 0\end{matrix}\right)\;,\;\;
\psi_{L}^{(n)}=\left(\begin{matrix}0,\cr\psiL^{(n)}\end{matrix}\right)\;,
\ee
and $\psiR$, $\psiL$ having two components.

\subsection{CS symmetry}

The (hermitian) $SU(2)_{CS}$ (shorter: CS for ``chiral spin'') algebra \cite{Glozman:2015qva} generators are
\be
CS_k:\quad T\in \{\gamma_k, \I \gamma_k \gamma_5, \gamma_5\}\;,\quad k=1,2,3,4\;.
\ee
We define the transformation guided by the Minkowski version (i.e., that $\psibar$ should transform like $\psidag\gamma_4$.)
\be
\psi \to\psi '= \E^{\I\alpha T}\psi \;,\quad 
 \psibar \to\psibar '= \psibar \gamma_4 \E^{-\I\alpha T^\dagger}\gamma_4
\ee
Then quark - antiquark bilinears transform like
\be\label{CS-bilin-transf}
\psibar\mathcal{O}\psi\to
\psibar  [\gamma_4\E^{-\I\alpha T_a^\dagger}\gamma_4 ]\mathcal{O} \E^{\I\alpha T_b}\psi 
\ee
where $T_a, T_b$ are generators of the algebra CS. Each operator this way is an element of  a multiplet. E.g., a CS$_1$ multiplet might be $(\gamma_2, \I\gamma_2\gamma_5,\I\gamma_4\gamma_3,\I\gamma_2\gamma_1)$ which corresponds to $\rho,\omega,a_1,f_1,b_1,h_1$ for isovectors and isoscalars.

For $T=\gamma_5$ this is the transformation of chiral symmetry \eq{chiralTrans}. Only $\gamma_5$ leaves the 
kinetic term invariant.  It turns out that only $\gamma_5$ is anomalous giving a factor for the Grassmann path 
integral integration measure. All  CS transformations leave the chemical potential term $\psibar \gamma_4\psi$ 
invariant. In \cite{Glozman:2014mka,Glozman:2015qva} the embedding of $SU(2)_{CS}\times SU(2)_f\subset 
SU(2 n_f)$ was suggested. In $SU(4)$ the vector mesons  form a 15-plet ($\rho, \rho', b_1,a_1.h_1,\omega, 
\omega'$) and a singlet ($f_1$).

The CS transformations as a whole are no symmetry of the Dirac action. However, it has been observed, that 
CS is a symmetry of certain meson and baryon masses, if the low lying (quasi-zero) modes are absent 
\cite{Denissenya:2014poa,Denissenya:2014ywa,Denissenya:2015mqa,Denissenya:2015woa}. At zero 
temperature, with artificial removal of low lying modes in the valence sector, confinement seems to persist. 
Above the chiral temperature  the zero modes are suppressed naturally. There are indications that some form 
of confinement persists as well \cite{Rohrhofer:2017grg}.
 
The chromo-electric observables $\psibar\gamma_4\psi$ are symmetric under CS, the kinetic term of the 
action and the chromo-magnetic terms $\psibar\gamma_k\psi$ ($k=1,2,3$) are not. Removing the near-zero 
modes apparently  restores the symmetry such that the influence of chromo-magnetism shrinks or disappears. 
One might conclude that confinement has its origin in the chromo-electric sector, which is symmetric under CS 
always \cite{Glozman:2017dfd}.

\section{Meson propagators}\label{sec_meson_props}
\begin{table}[b]
\begin{center}
\begin{tabular}{|rr|rr|r|}
\hline\hline
$\Gamma_{src}$ &$\Gamma_{snk}$         &$\I\Gamma_{src}\gamma_5$ &$\I\gamma_5\Gamma_{snk}$                    &$s_5$\\
 \hline
$\mathbf{1}$       & $  \mathbf{1}$ 		& $\I\,\gamma_5$          		  & $\I\,\gamma_5$ 			& $1 $ \\  
$\gamma_k $     & $   \gamma_k$  		& $\I\,\gamma_k \gamma_5$   & $-\I\,\gamma_k \gamma_5$ &$-1$ \\  
$\gamma_4$      & $  \gamma_4$ 		& $\I\,\gamma_4 \gamma_5$   & $-\I\,\gamma_4 \gamma_5$ 	&$-1$\\  
$\gamma_k \gamma_j$ & $  -\gamma_k \gamma_j$  & $\I\,\gamma_k \gamma_j \gamma_5$   & $ - \I\,\gamma_k \gamma_j \gamma_5$ &$1$  \\  
$\gamma_k \gamma_4$& $  -\gamma_k \gamma_4$  & $\I\,\gamma_k \gamma_4 \gamma_5$   & $ - \I\,\gamma_k \gamma_4 \gamma_5$ &$1$  \\  
\hline
\end{tabular}
\end{center}
\caption{\label{sosi_gammas}We list the sink and source operator kernels.
We also give sign factors $s_5$ defined by  $\Gamma \gamma_5=s_5 \gamma_5 \Gamma$.
}
\end{table}

We restrict ourselves to two mass degenerate quark flavors $u$ and $d$.  As already mentioned, we neglect exact zero modes, either because we are in that topological sector or  because they have been removed. 

We study the propagators for mesons of type $\overline{\Psi}(\vec{\tau}\otimes \Gamma)\Psi$ and $\overline{\Psi}(\mathbf{1}_f\otimes \Gamma)\Psi$. The $\Gamma$ are listed in Table \ref{sosi_gammas}; the choice has been motivated by the discussion of the CS symmetry in Ref. \cite{Glozman:2015qva}. We emphasize that our results are for the propagators themselves, not just the masses.

For the isotriplet we introduce the connected propagator for a given background gauge field $A$,
\bea
P_{c}(\Gamma,x,y)&=&-[\ubar_{xa} \Gamma^{snk}_{ab} d_{xb}\dbar_{yc} \Gamma^{src}_{cd} u_{yd}]_A\nonumber\\
&=&
\tr_{Dirac}[{D_u}^{-1}_{yd|xa} \Gamma^{snk}_{ab}{D_d}^{-1}_{xb|yc} \Gamma^{src}_{cd} ]
\eea
where $[\ldots ]_A$ indicates Grassmann integration in an external field $A$. We relate source and sink by 
$\Gamma_{snk}=\Gamma_{src}^\dagger\equiv s_\Gamma\Gamma_{src}$, see Table \ref{sosi_gammas}. For degenerate quark masses $D_u^{-1}=D_d^{-1}$. With the spectral representation for $D^{-1}$ the meson propagator becomes
\bea
P_{c}(\Gamma,x,y)&=&s_\Gamma \sum_{n,k} \frac{1}{(\eta_n+\I \,m)(\eta_k+\I \,m)}\\
&&\qquad 
\psi^{(n)}_{yd}\psi^{(n)\dagger}_{xa}
\Gamma_{ab} \psi^{(k)}_{xb}\psi^{(k)\dagger}_{yc}\Gamma_{cd}\nonumber\\
&=&s_\Gamma\sum_{n,k} \frac{1}{(\eta_n+\I \,m)(\eta_k+\I \,m)}\nonumber\\
&&\qquad (\psi^{(n)\dagger}_{xa}\Gamma_{ab} \psi^{(k)}_{xb})
(\psi^{(k)\dagger}_{yc}\Gamma_{cd}\psi^{(n)}_{yd})\;.\nonumber
\eea
The isoscalar propagators have also disconnected contributions proportional to
\be
P_{d}(\Gamma,x)=s_\Gamma
\tr_{Dirac}[{D_u^{-1}}_{xb|xa} \Gamma_{ab}]\;\tr_{Dirac}[{D_d}^{-1}_{yd|yc} \Gamma_{cd}]
\ee
Like other expressions used here, this has to be regularized (e.g., by lattice regularization) and there 
are standard tools to do this (e.g.,  \cite{Fujikawa:1979ay,Fujikawa:1980eg,Fujikawa:2004}).
The results and conclusions presented here are not affected.

We will find useful the identities
\bea\label{identities}
\psi^{(-n)\dagger}_{xa}\Gamma_{ab}  \psi^{(-k)}_{xb}&=&
\psi^{(n)\dagger}_{xa}(\gamma_5 \Gamma\gamma_5)_{ab}  \psi^{(k)}_{xb}\nonumber\\
&=&s_{5} \psi^{(n)\dagger}_{xa} \Gamma_{ab}\psi^{(k)}_{xb}\\
\psi^{(-n)\dagger}_{xa}\Gamma_{ab}  \psi^{(k)}_{xb}&=&
\psi^{(n)\dagger}_{xa} (\gamma_5\Gamma)_{ab} \psi^{(k)}_{xb}\nonumber\\
&=&s_{5} \psi^{(n)\dagger}_{x,a} \Gamma_{ab}  \psi^{(-k)}_{x,b}\;,
\eea
where $\gamma_5 \Gamma\gamma_5=s_{5} \Gamma$ (see Table \ref{sosi_gammas}).

\begin{figure}[tb]
\centering{
\includegraphics[angle=0,width=0.45\linewidth]{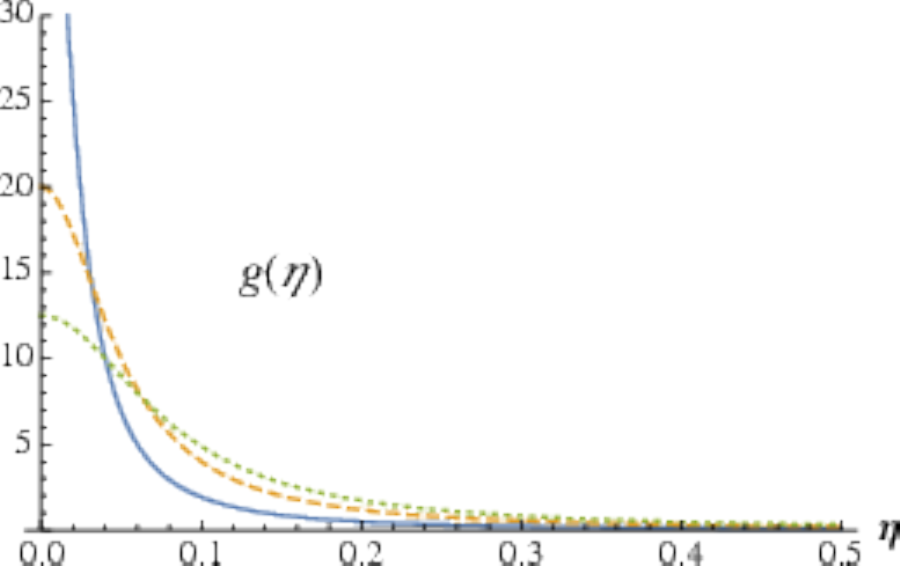}
\includegraphics[angle=0,width=0.45\linewidth]{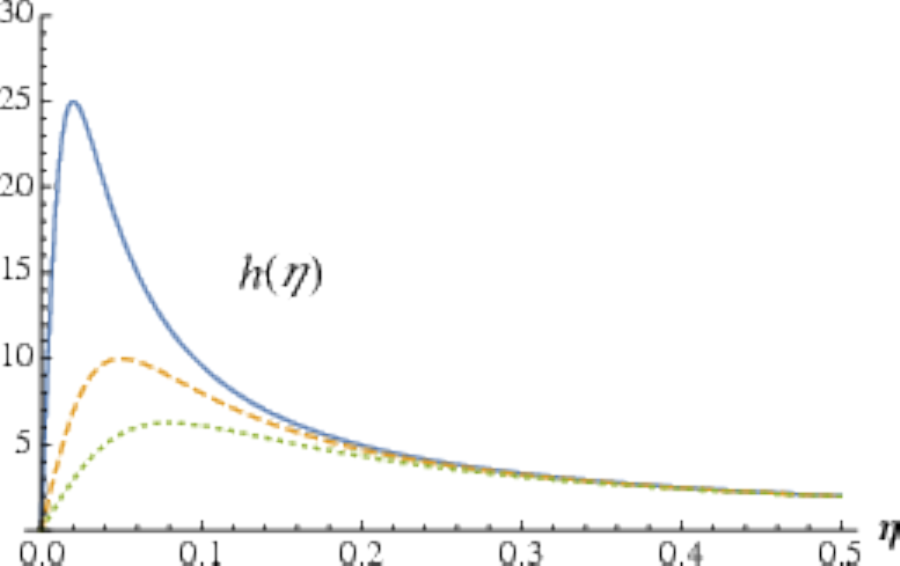}
}
\caption{\label{fig_g_h} The weight functions $g(\eta)$ and $h(\eta)$  from 
\eq{eq_g_h}  for typical values $m$ equal to 0.02 (full), 
0.06 (dashed) and 0.08 (dotted).}
\end{figure}

The difference between two meson propagators depends on 
\begin{itemize}
\item the generic distribution density of eigenvalues $\rho(m,\eta)$ (which depends on the gauge configuration and the Dirac operator),
\item the values of the overlap matrix elements $\psidag\Gamma\psi$ (which are bounded from above due to the orthogonality and normalization of the eigenvectors), and 
\item a weight function discussed below. 
\end{itemize}
The generic distribution of eigenvalues $\rho(m,\eta)$ is needed only for the small eigenvalues, where it  vanishes fast enough or  there is even a gap, the relevant cases of this study. The bulk behavior is inconspicuous \cite{Verbaarschot:1997bf}. 

We focus here on the third factor. As derived in the Appendix
the functions (see Fig.s \ref{fig_g_h}) 
\be\label{eq_g_h}
g(m,\eta)\equiv  \frac{m }{m^2 + \eta^2} \;,\quad
h(m,\eta)\equiv \frac{\eta}{m^2 + \eta^2} \;,
\quad (\eta > 0)
\ee
turn up in the sums over eigenvalues in the next section.  They are essential for the argumentation. Both functions give large weight to contributions from small $\eta$. The function $g$ is peaked at $\eta=0$ and approaches $(\pi/2)\delta(\eta)$  for small masses $m\to 0$; for large $\eta$ it falls like $1/\eta^2$. Propagator differences weighted by $g$ vanish for small $m$ if there is a gap in the density $\rho(m,\eta)$ at low lying eigenmodes, i.e.  if there are no eigenvalues below some value, or if the density vanishes fast enough for $\eta\to 0$. Propagators that differ only by terms proportional to $g$ will be called {\em $g$-equivalent}.

The function $h$ is peaked at $\eta=m$ and falls like $1/\eta$ for large $\eta$. Compared to $g$ this behaviour may not suppress the higher modes enough, depending on the Dirac structure. Propagators that differ also by terms proportional to $h$ will be called {\em $h$-equivalent}. For these the existence of a gap at low eigenvalues is not sufficient to  obtain propagator agreement and more conditions have to be met.

In the next section we  discuss the main results.The full derivations can be found in the appendix.
The resulting equivalences between the meson propagators are listed in Table \ref{table_relations} and 
 shown in Fig.s \ref{fig_relations_vector} and  \ref{fig_relations_scalar}. 
   
\subsection{$g$-equivalent mesons}

\subsubsection{$\Gamma$ vs. $\I\Gamma\gamma_5$}\label{sec_conn_g_g5}

For  $\Gamma\in\{\mathbf{1},$ $\gamma_k,$ $\gamma_4,$ $\gamma_k\gamma_j, $ $\gamma_k\gamma_4 \}$
the difference between meson isovector propagators is 
\bea\label{case1}
P_{c}(\Gamma,x,y)-P_{c}(\I\,\Gamma\gamma_5,x,y)=&&\\
&&\hspace{-25mm}-8\sum_{n>0,k>0}g(m,\eta_n)g(m,\eta_k)\nonumber\\
&&\hspace{-15mm}\big[\psi^{(k)\dagger}_{x,a}\Gamma_{a,b}\psi^{(n)}_{x,b}\psi^{(n)\dagger}_{y,c}\Gamma_{c,d}  \psi^{(k)}_{y,d} \nonumber\\
&&\hspace{-15mm}+\psi^{(k)\dagger}_{x,a}\Gamma_{a,b}\psi^{(-n)}_{x,b}\psi^{(-n)\dagger}_{y,c}\Gamma_{c,d}  \psi^{(k)}_{y,d}\big]\;.\nonumber
\eea
These propagator pairs are $g$-equivalent.

For small masses $m\to 0$ the functions $g$  emphasize the  contributions of small eigenvalues. If the eigenvalue density $\rho(m,\eta)$  vanishes at small eigenvalues, then the propagator difference vanishes as well and axial symmetry is restored. The factors with eigenvectors are bounded (the eigenvectors are normalized).
 
The integral over $x, y$ and sum over $a$ and the other, hidden indices gives the susceptibility. For $\Gamma=\mathbf{1}$ the second  term in \eq{case1} vanishes due to orthogonality. The first term gives $\delta_{nk}$. The susceptibility difference (the $U_A(1)$ susceptibility) then is
\bea\label{susc_case1}
\chi(\mathbf{1})-\chi(\I\, \gamma_5)&=&- \frac{4}{V}\sum_{n>0}g(m,\eta_n)^2\nonumber\\
&\simeq&\int_0^\infty d\eta\;\rho(m,\eta)g(m,\eta)^2.
\eea
for the eigenvalue density $\rho(m,\eta)$ (cf. \cite{Cohen:1996ng}, the discussion in \cite{Fukaya:2017wfq}
and \cite{Ohno:2012br,Aoki:2012yj,Kovacs:2008sc,Tomiya:2014mma,Dick:2015twa}). 
This term vanishes if there is a gap in  $\rho(m,\eta)$ or if the density vanishes fast enough\footnote{A behavior $\lim_{m\to 0}\rho(m,\eta)=\mathcal{O}(\eta^3)$ is sufficient for the vanishing of the $U_A(1)$ susceptibility 
\cite{Aoki:2012yj}.} for $\eta\to 0$.
The susceptibilities for the other difference pairs of Table \ref{sosi_gammas} are also $g$-equivalent.

\subsubsection{Isovector vs. Isoscalar}\label{sec_isoscalar}

Isoscalar propagators have also disconnected contributions. For $\Gamma\in\{\mathbf{1},$ $\gamma_k\gamma_j$, $\gamma_k\gamma_4$, $\I\,\gamma_5$, $\I\,\gamma_k\gamma_j\gamma_5$, $ \I\,\gamma_k\gamma_4\gamma_5 \}$ they
 have the form (for the derivation see Appendix \ref{app_b})
\bea\label{case2result}
&&\hspace{-8mm}\sum_{n>0,k>0}
4\,g(m,\eta_n) g(m,\eta_k) \nonumber\\
 && \times(\psi^{(n)\dagger}(x)\Gamma_{snk}\psi^{(n)}(x))( \psi^{(k)\dagger}(y)\Gamma_{src}\psi^{(k)}(y)\,.
\eea
We find that if the eigenvalue density at small eigenvalues vanishes,  the disconnected term vanishes as well. The isoscalar is $g$-equivalent to the isovector propagators for these $\Gamma$. Combining the relations with those of Sect. \ref{sec_conn_g_g5} one obtains further relations between {\em isoscalar} pairs
$(\mathbf{1},\gamma_5)$, $(\I\gamma_k\gamma_j,\I\gamma_k\gamma_j\gamma_5)$ and $(\gamma_k\gamma_4,\gamma_k\gamma_4\gamma_5)$.
The scalar mesons at high temperature were studied in \cite{Tomiya:2016jwr}.

\subsection{$h$-equivalent mesons}

\subsubsection{More disconnected terms} \label{sec_more_disc_terms}

Compared with $g$-equivalence the now discussed type is more subtle with factors $h(m,\eta)$, needing further bounds or eigenmode properties to find propagator agreement. To see this we use the chiral basis of Sect. \ref{sec_chiral_symmetry}.

The disconnected contributions for propagators with $\Gamma\in\{\gamma_k,\gamma_4,\gamma_k\gamma_5, \gamma_4\gamma_5 \}$
have terms with factors like
\bea\label{disconn_g_g5_h_type}
&&\hspace{-8mm}
h(m,\eta_n) h(m,\eta_k) \nonumber\\
&&\times(\psiR^{(n)\dagger}(x)\,\sigma \,\psiL^{(n)}(x))
(\psiL^{(k)\dagger}(y)\,\sigma \,\psiR^{(k)}(y))\,,
\eea
where $\sigma$ is a $2\times 2$ matrix (i.e., a sub-block of $\Gamma$; for the complete expression see App. \ref{app_c}). The prefactor again favors low eigenmodes for small $m$. Therefore the disconnected contributions become much smaller if the low modes are suppressed in the generic density.

Even if the low modes are absent, however,  the higher modes still contribute to the difference more than in the $g$-equivalent case since $h(m,\eta)$ decreases slower with $\eta$ than $g(m,\eta)$. The quality of the agreement then depends on the matrix elements in \eq{disconn_g_g5_h_type} and not only on the eigenvalue density. This is discussed in the subsequent section.

If the high modes contribution can be neglected the isoscalar propagator agrees with the isovector propagator for the listed $\Gamma$. Considering the results for the connected propagators  this implies also agreement of the isoscalar propagator pairs $(\gamma_k,\gamma_k\gamma_5)$ and $(\gamma_4,\gamma_4\gamma_5)$. 

\subsubsection{ $\Gamma$ vs. $\Gamma\gamma_4$}

Finally let us  consider the connected propagator pairs for $(\Gamma,\Gamma\gamma_4)$ for $\Gamma\in\{\mathbf{1},\gamma_k,
\gamma_5,  \gamma_k\gamma_j,  \gamma_k\gamma_5 \}$; these are central for the CS symmetry.
The propagator differences are sums of two types of terms
\bea
&&\hspace{-8mm}g(m,\eta_n)g(m,\eta_k)\nonumber\\
&&
\big(
(\psiL^{(n)\dagger}(x)\,\sigma \,\psiL^{(k)}(x))
(\psiL^{(k)\dagger}(y)\,\sigma \,\psiL^{(n)}(y))\nonumber\\
&&+
 (\psiR^{(n)\dagger}(x)\,\sigma \,\psiR^{(k)}(x))
 (\psiR^{(k)\dagger}(y)\,\sigma \,\psiR^{(n)}(y))\nonumber\\
&&  -
 (\psiR^{(n)\dagger}(x)\,\sigma \,\psiL^{(k)}(x))
 (\psiL^{(k)\dagger}(y)\,\sigma \,\psiR^{(n)}(y))\nonumber\\
 && -
 (\psiL^{(n)\dagger}(x)\,\sigma \,\psiR^{(k)}(x))
 (\psiR^{(k)\dagger}(y)\,\sigma \,\psiL^{(n)}(y))\big)
 \eea
and 
\bea
&&\hspace{-8mm}h(m,\eta_n)h(m,\eta_k)\nonumber\\
&&\big(
(\psiR^{(n)\dagger}(x)\,\sigma \,\psiR^{(k)}(x))
 (\psiL^{(k)\dagger}(y)\,\sigma \,\psiL^{(n)}(y))\nonumber\\
 && +
 (\psiL^{(n)\dagger}(x)\,\sigma \,\psiL^{(k)}(x))
 (\psiR^{(k)\dagger}(y)\,\sigma \,\psiR^{(n)}(y))\nonumber\\
 && -
( \psiL^{(n)\dagger}(x)\,\sigma \,\psiR^{(k)}(x))
( \psiL^{(k)\dagger}(y)\,\sigma \,\psiR^{(n)}(y))\nonumber\\
 &&-
 (\psiR^{(n)\dagger}(x)\,\sigma \,\psiL^{(k)}(x))
 (\psiR^{(k)\dagger}(y)\,\sigma \,\psiL^{(n)}(y))\big)
 \eea
The first term  becomes negligible if the fermion mass is small and if there is a gap in $\rho(m,\eta)$ at small $\eta$. 
In the second term, unlike the connected propagators discussed in Sect. \ref{sec_conn_g_g5}, 
all four types $\psiR^{\dagger}\,\sigma \,\psiL$, $\psiL^{\dagger}\,\sigma \,\psiR$, $\psiR^{\dagger}\,\sigma \,\psiR$,
$\psiL^{\dagger}\,\sigma \,\psiL$ enter the propagator difference multiplying $h$.

\begin{table}[tb]
\begin{center}
\begin{tabular}{rcl}
\hline
\multicolumn{3}{c}{$g$-equivalent meson propagators}\\
\hline
$\tau^a\otimes\mathbf{1}$&$\leftrightarrow$&$\tau^a\otimes\gamma_5$\\
$\tau^a\otimes\gamma_k$         &$\leftrightarrow$&  $\tau^a\otimes\gamma_k\gamma_5$          \\
$\tau^a\otimes\gamma_4$         &$\leftrightarrow$&  $\tau^a\otimes\gamma_4\gamma_5$          \\
$\tau^a\otimes\gamma_k\gamma_j$ &$\leftrightarrow$&  $\tau^a\otimes\gamma_k\gamma_j\gamma_5$  \\
$\tau^a\otimes\gamma_k\gamma_4$ &$\leftrightarrow$&  $\tau^a\otimes\gamma_k\gamma_4\gamma_5$  \\
$\tau^a\otimes\mathbf{1}$&$\leftrightarrow$&$\mathbf{1}_f\otimes\mathbf{1}$\\    
$\tau^a\otimes\gamma_k\gamma_j$ &$\leftrightarrow$&$\mathbf{1}_f\otimes\gamma_k\gamma_j$ \\   
$\tau^a\otimes\gamma_k\gamma_4$ &$\leftrightarrow$&$\mathbf{1}_f\otimes\gamma_k\gamma_4$ \\  
$\tau^a\otimes\gamma_5$ &$\leftrightarrow$&$\mathbf{1}_f\otimes\gamma_5$ \\ 
$\tau^a\otimes\gamma_k\gamma_j\gamma_5$&$\leftrightarrow$&$\mathbf{1}_f\otimes\gamma_k\gamma_j\gamma_5$\\
$\tau^a\otimes\gamma_k\gamma_4\gamma_5$&$\leftrightarrow$&$\mathbf{1}_f\otimes\gamma_k\gamma_4\gamma_5$\\
$\mathbf{1}_f\otimes\mathbf{1}$ &$\leftrightarrow$&$\mathbf{1}_f\otimes\gamma_5$ \\ 
$\mathbf{1}_f\otimes\gamma_k\gamma_j$&$\leftrightarrow$&$\mathbf{1}_f\otimes\gamma_k\gamma_j\gamma_5$\\
$\mathbf{1}_f\otimes\gamma_k\gamma_4$&$\leftrightarrow$&$\mathbf{1}_f\otimes\gamma_k\gamma_4\gamma_5$\\
\hline
\multicolumn{3}{c}{$h$-equivalent meson propagators}\\
\hline
$\tau^a\otimes\gamma_k$&$\leftrightarrow$&$ \mathbf{1}_f\otimes\gamma_k$\\          
$\tau^a\otimes\gamma_4$&$\leftrightarrow$&$\mathbf{1}_f\otimes\gamma_4$          \\
$\tau^a\otimes\gamma_k\gamma_5$&$\leftrightarrow$&$\mathbf{1}_f\otimes\gamma_k\gamma_5$\\  
$\tau^a\otimes \gamma_4\gamma_5$&$\leftrightarrow$&$\mathbf{1}_f\otimes \gamma_4\gamma_5$\\ 
$\tau^a\otimes \mathbf{1}$&$\leftrightarrow$&$         \tau^a\otimes \gamma_4$\\
$\tau^a\otimes \gamma_k$&$\leftrightarrow$&$\tau^a\otimes \gamma_k\gamma_4$\\
$\tau^a\otimes \gamma_5$&$\leftrightarrow$&$\tau^a\otimes \gamma_4\gamma_5$\\
$\tau^a\otimes \gamma_k\gamma_j$&$\leftrightarrow$&$\tau^a\otimes \gamma_k\gamma_j\gamma_4$\\
$\tau^a\otimes \gamma_k\gamma_5$&$\leftrightarrow$&$\tau^a\otimes \gamma_k\gamma_4\gamma_5$\\
$ \mathbf{1}_f\otimes\gamma_k$&$\leftrightarrow$&$ \mathbf{1}_f\otimes\gamma_k\gamma_5$\\          
$ \mathbf{1}_f\otimes\gamma_4$&$\leftrightarrow$&$\mathbf{1}_f\otimes\gamma_4\gamma_5$          \\
$\mathbf{1}_f\otimes \mathbf{1}$&$\leftrightarrow$&$         \mathbf{1}_f\otimes \gamma_4$\\
$\mathbf{1}_f\otimes \gamma_k$&$\leftrightarrow$&$\mathbf{1}_f\otimes \gamma_k\gamma_4$\\
$\mathbf{1}_f\otimes \gamma_5$&$\leftrightarrow$&$\mathbf{1}_f\otimes \gamma_4\gamma_5$\\
$\mathbf{1}_f\otimes \gamma_k\gamma_j$&$\leftrightarrow$&$\mathbf{1}_f\otimes \gamma_k\gamma_j\gamma_4$\\
$\mathbf{1}_f\otimes \gamma_k\gamma_5$&$\leftrightarrow$&$\mathbf{1}_f\otimes \gamma_k\gamma_4\gamma_5$\\
\hline
\end{tabular}
\end{center}
\caption{\label{table_relations}Related meson propagators; For $g$-equivalent propagators the differences 
vanish  in the massless limit if there are no low lying modes. Further assumptions are necessary for $h$-equivalence.}
\end{table}

When there are no eigenvalues below some $|\eta|<\eta_0$ or the generic density vanishes fast enough towards $\eta=0$ the propagator difference is dominated  by the terms with $h$. The factors $\psiL^{\dagger}\,\sigma \,\psiR$, etc., encode the dynamics of QCD. 

\begin{figure}[tb]
\centering{
\includegraphics[angle=0,width=0.9\linewidth,clip]{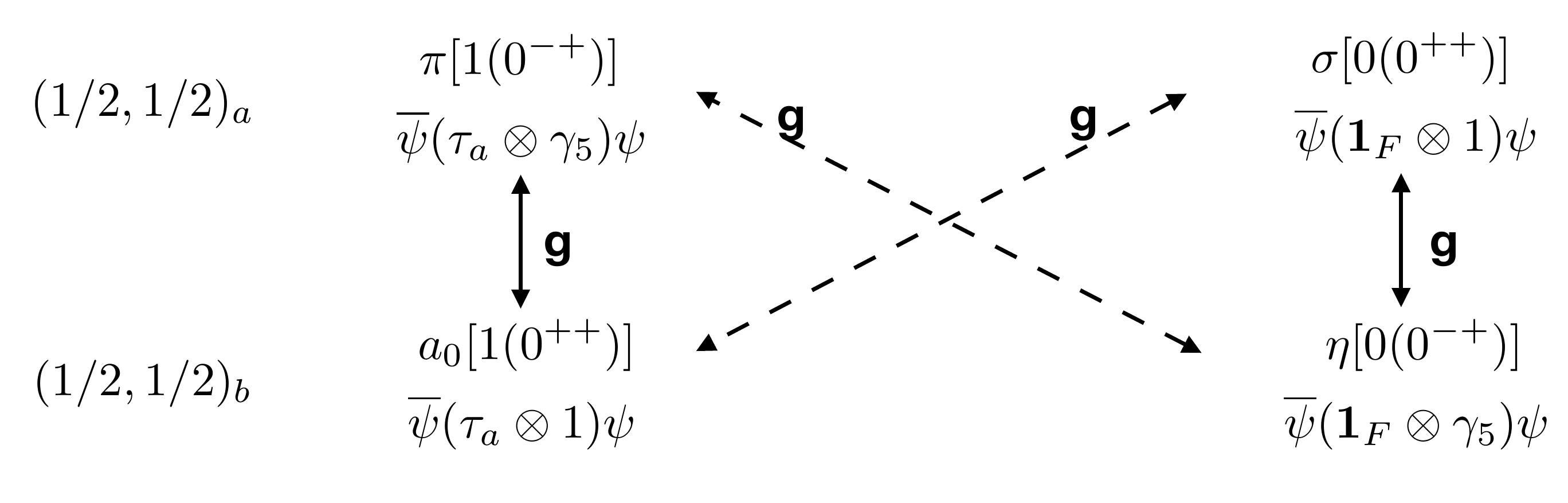}
}
\caption{\label{fig_relations_scalar}The equivalence relations between the corresponding meson propagators for scalar mesons are shown.  The $h$-equivalences have been omitted (although they are listed in Table \ref{table_relations})  as they relate to unphysical states, The left-hand column indicates the chiral structure \cite{Cohen:1996sb}. }
\end{figure}

There are a few observations that may shed some light:
\begin{itemize}
\item[-]
The $h$- and $g$-terms of the difference pairs $(\mathbf{1}, \gamma_4)$ and $(\I\gamma_5, \I\gamma_4\gamma_5)$ are identical, as are those for $(\gamma_k,\gamma_k\gamma_4)$ and 
$(\I\gamma_k\gamma_5, \I\gamma_k\gamma_4\gamma_5)$. In other words, if the propagator for $\gamma_k$ and  $\gamma_k\gamma_4$  agree, so do the propagators for $\I\gamma_k\gamma_5$ and $\I\gamma_k\gamma_4\gamma_5$.
\item[-]    
The $h$-terms vanish for chiral eigenmodes  of the form  $(\psiR,0)$ or $(0,\psiL)$ or will be suppressed for almost chiral eigenmodes (where either $|\psiR| \gg |\psiL|$ or $|\psiR| \ll |\psiL|$). However, such behaviour is expected mainly for the low lying modes which are truncated or suppressed anyhow in the situation of relevance here.
\item[-]
The mesons with $\Gamma=$ $\gamma_4$ or $\gamma_4\gamma_5$ have only terms $\psiR^{\dagger}\psiL$ 
and $\psiL^{\dagger}\psiR$; now  $\psiR^{\dagger}$ and $\psiL$ correspond to the same 
helicity which cannot add up to zero. The states cannot be physical scalars  \cite{Glozman:2015qva}. For this reason we omit these states in Fig. \ref{fig_relations_scalar}.
\item[-]
There is numerical evidence \cite{Denissenya:2014ywa} indicating that the scalar propagators show less agreement than the vector propagators. This is a hint  that the vector matrix elements $\psi^{(n)\dagger} \gamma_j \psi^{(k)}$ are smaller than the scalar ones  $\psi^{(n)\dagger}  \psi^{(k)}$.
\end{itemize}

\begin{figure}[tb]
\centering{
\includegraphics[angle=0,width=0.9\linewidth,clip]{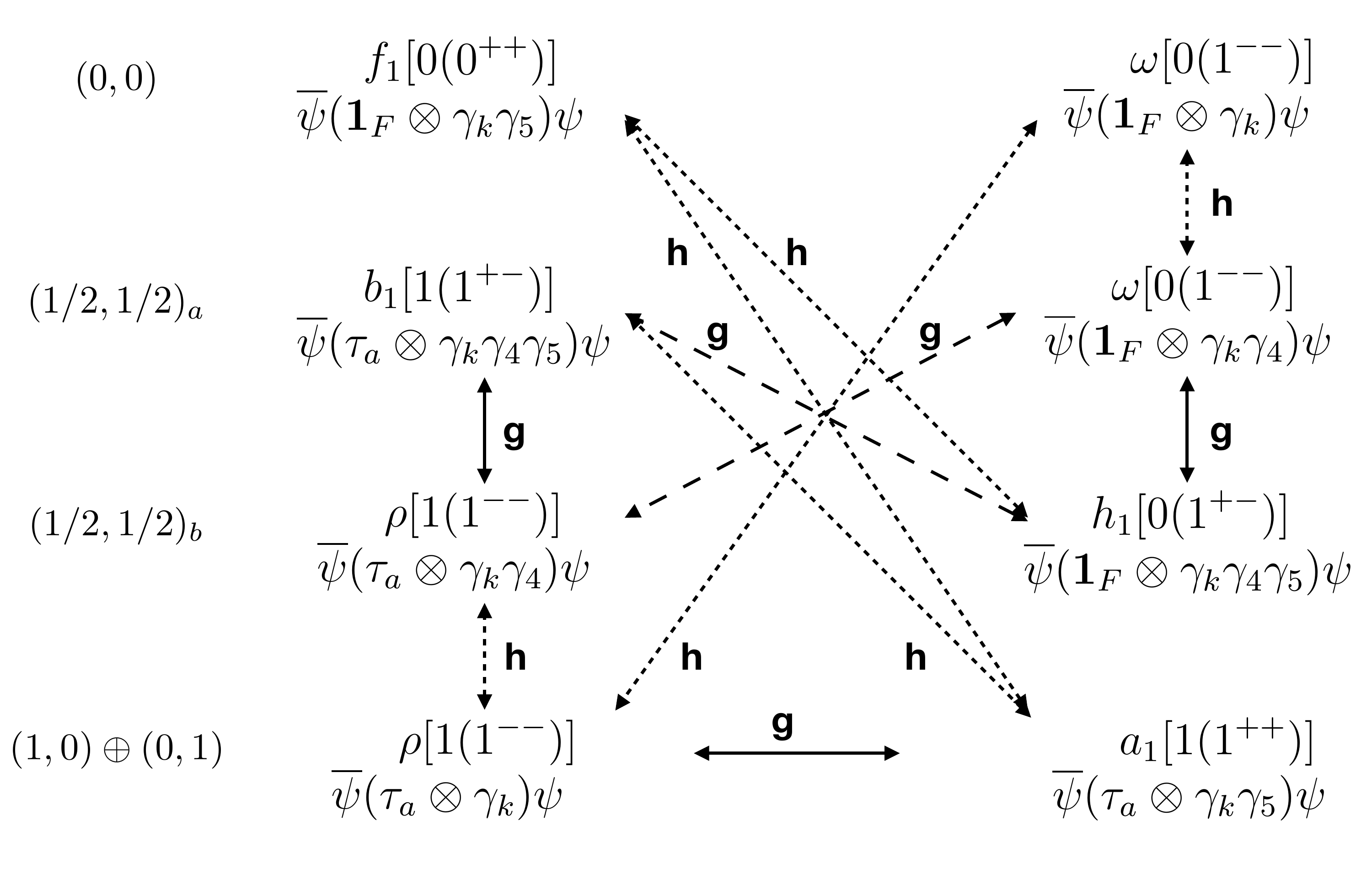}
}
\caption{\label{fig_relations_vector}The equivalence relations between the corresponding meson propagators for vector mesons are shown. The arrows symbolize the entries in Table \ref{table_relations} and the equivalence  type  as discussed in the text is shown. The arrangement of
the operators follows \cite{Glozman:2015qva} for better comparison; the left-hand column indicates the chiral structure \cite{Cohen:1996sb}. }
\end{figure}

\section{Conclusions}
 
Here  we studied the r\^ole of low lying eigenmodes of the Dirac operator in meson propagators.  The study is motivated by lattice QCD calculations where it was found that the differences between meson propagators of a large class disappear if the low lying  (i.e., close to zero) modes of the Dirac operator are suppressed. The mass  degeneracies have been observed when the low modes were truncated explicitly 
\cite{Denissenya:2014poa,Denissenya:2014ywa,Denissenya:2015mqa,Glozman:2015qva}
or dynamically suppressed at large temperature \cite{Rohrhofer:2017grg}.\footnote{A dominance 
of the low lying modes for some meson propagators has been noticed earlier (see, e.g., 
\cite{DeGrand:2000gq}).}

There are two  qualitatively different kinds of relations. Those with a weight factor $g(m,\eta)$  we call $g$-equivalent. Meson propagators that are $g$-equivalent (Fig.s \ref{fig_relations_vector} and \ref{fig_relations_scalar} and table \ref{table_relations}) approach each other for small quarkmass, if there is a low-eigenvalue suppression or gap in the generic eigenvalue density. These equivalences, when realized, restore the axial symmetries $SU(2)_A$ and $U(1)_A$.

The second type called $h$-equivalence needs further constraints in order to provide vanishing propagator differences. The  weight factor  $h(m,\eta)$ is also peaked at small $\eta$ but does not  suppress the higher modes as efficient. In that case the quality of agreement depends on the overlap of eigenvectors.

We find:
\begin{itemize}
\item The connected (isovector) propagators $P_c(\Gamma)$ and $P_c(\I\Gamma\gamma_5)$ for
$\Gamma\in\{\mathbf{1},$ $\gamma_k,$ $\gamma_4,$ $\gamma_k\gamma_j, $ $\gamma_k\gamma_4 \}$  differ only by $g$-type terms.  If there is a low mode suppression the propagators of a pair  agree with each other for $m\to 0$. The susceptibilities of the connected (isovector) propagators $P_c(\Gamma)$ inherit the $g$-equivalence property.
\item For some isoscalar mesons (see Sect. \ref{sec_isoscalar}) the propagators'  disconnected  contributions are $g$-type terms. For these mesons the isoscalar and isovector propagators agree in the massless limit if there is a suppression of low eigenvalues.
\item The connected (isovector) propagators $P_c(\Gamma)$ and $P_c(\Gamma\gamma_4)$ for
$\Gamma\in\{\mathbf{1}$, $\gamma_k$, $\gamma_5$, $ \gamma_k\gamma_j$, $ \gamma_k\gamma_5 \}$. differ by $g$-type and $h$-type terms. The $h$-terms become small for almost chiral eigenmodes (where either  $|\psiR| \gg |\psiL|$ or $|\psiR| \ll |\psiL|$) or small overlap  $\phi^{(n)\dagger}\Gamma\phi^{(k)}$. 
\item The propagator difference  ($P_c(\gamma_k )-P_c(\gamma_k\gamma_4)$) differs from ($P_c(\I\gamma_k\gamma_5 )-P_c(\I\gamma_k\gamma_4\gamma_5)$) only by $g$-type terms.  I.e., if the $h$-type contribution vanishes for one pair it also vanishes for the other. Also the propagator difference  ($P_c(\mathbf{1} )-P_c(\gamma_4)$) differs from ($P_c(\I\gamma_5 )-P_c(\I\gamma_4\gamma_5)$) only by $g$-type terms.
\end{itemize}
In summary the axial symmetries between the meson propagators and susceptibilities are recovered for decreasing quark mass upon suppression of low lying eigenmodes in the eigenvalue distribution. A similar behaviour for the observed $\gamma_4$ symmetry requires in addition small overlap of the higher eigenmodes.

The emerging agreement between the meson propagators  explains numerical lattice QCD results for meson mass degeneracies. Based on the meson mass pattern the symmetries CS and $SU(4) $ were conjectured \cite{Glozman:2015qva}. These may have far-reaching consequences  \cite{Glozman:2017dfd}.

\acknowledgements%
{I profited much from discussions with Christof Gattringer and Vasily Sazonov.
Many thanks go to Leonid Glozman for numerous discussions,  for reading the manuscript, and for his persistence. }

\appendix
\section{$P_c(\Gamma,x,y)-P_c(\Gamma\gamma_5,x,y)$}\label{app_a}

The connected propagator $P_c(\Gamma)$ for $\Gamma$ is written in terms of the spectral representation of the quark propagators:
\be 
P_c(\Gamma)=s_\Gamma\sum_{n,k }f_n f_k  \psi^{(k)\dagger}_{x,a}\Gamma_{a,b}\psi^{(n)}_{x,b}\psi^{(n)\dagger}_{y,c}\Gamma_{c,d}  \psi^{(k)}_{y,d}\;,
\ee
with   $s_\Gamma$ defined in Sect.  \ref{sec_meson_props} and Table \ref{sosi_gammas} and summation over paired indices is implied. We use the abbreviation
\be 
f_n\equiv\frac{1}{\eta_n+\I \,m}
\ee
with $\eta_{-n}=-\eta_n$. There are no exact zero modes by assumption. We rewrite the sum like
\bea\label{propG}
&s_\Gamma\sum_{n>0,k>0}&\Big[ 
 f_n f_k \psi^{(k)\dagger}_{x,a}\Gamma_{a,b}\psi^{(n)}_{x,b}\psi^{(n)\dagger}_{y,c}\Gamma_{c,d}  \psi^{(k)}_{y,d}\nonumber\\
&&\hspace{-10mm}+f_{-n} f_{-k} \psi^{(-k)\dagger}_{x,a}\Gamma_{a,b}\psi^{(-n)}_{x,b}\psi^{(-n)\dagger}_{y,c}\Gamma_{c,d}  \psi^{(-k)}_{y,d}\nonumber\\
&&\hspace{-10mm}+f_{-n} f_k \psi^{(k)\dagger}_{x,a}\Gamma_{a,b}\psi^{(-n)}_{x,b}\psi^{(-n)\dagger}_{y,c}\Gamma_{c,d}  \psi^{(k)}_{y,d}\nonumber\\
&&\hspace{-10mm}+f_n f_{-k} \psi^{(-k)\dagger}_{x,a}\Gamma_{a,b}\psi^{(n)}_{x,b}\psi^{(n)\dagger}_{y,c}\Gamma_{c,d}  \psi^{(-k)}_{y,d}\Big]\nonumber\\
&=s_\Gamma\sum_{n>0,k>0}&\Big[ 
f_n f_k \psi^{(k)\dagger}_{x,a}\Gamma_{a,b}\psi^{(n)}_{x,b}\psi^{(n)\dagger}_{y,c}\Gamma_{c,d}  \psi^{(k)}_{y,d}\nonumber\\
&&\hspace{-10mm}+f_{-n} f_{-k} \psi^{(k)\dagger}_{x,a}\Gamma_{a,b}\psi^{(n)}_{x,b}\psi^{(n)\dagger}_{y,c}\Gamma_{c,d}  \psi^{(k)}_{y,d}\nonumber\\
&&\hspace{-10mm}+f_{-n} f_k \psi^{(k)\dagger}_{x,a}\Gamma_{a,b}\psi^{(-n)}_{x,b}\psi^{(-n)\dagger}_{y,c}\Gamma_{c,d}  \psi^{(k)}_{y,d}\nonumber\\
&&\hspace{-10mm}+f_n f_{-k} \psi^{(k)\dagger}_{x,a}\Gamma_{a,b}\psi^{(-n)}_{x,b}\psi^{(-n)\dagger}_{y,c}\Gamma_{c,d}  \psi^{(k)}_{y,d}\Big]\nonumber\\
&=s_\Gamma\sum_{n>0,k>0}&\Big[ 
(f_n f_k +f_{-n} f_{-k})\nonumber\\
&&\hspace{-10mm} \psi^{(k)\dagger}_{x,a}\Gamma_{a,b}\psi^{(n)}_{x,b}\psi^{(n)\dagger}_{y,c}\Gamma_{c,d}  \psi^{(k)}_{y,d}\nonumber\\
&&\hspace{-10mm}+(f_{-n} f_k+f_n f_{-k}) \nonumber\\
&&\hspace{-10mm}\psi^{(k)\dagger}_{x,a}\Gamma_{a,b}\psi^{(-n)}_{x,b}\psi^{(-n)\dagger}_{y,c}\Gamma_{c,d}  \psi^{(k)}_{y,d}\Big]
 \nonumber\\
&=s_\Gamma\sum_{n>0,k>0}&\Big[ 
(-2 g(m,\eta_n) g(m,\eta_k) + 2 h(m,\eta_n)h(m,\eta_k))\nonumber\\
&&\hspace{-10mm} \psi^{(k)\dagger}_{x,a}\Gamma_{a,b}\psi^{(n)}_{x,b}\psi^{(n)\dagger}_{y,c}\Gamma_{c,d}  \psi^{(k)}_{y,d}\nonumber\\
&&\hspace{-10mm}+(-2 g(m,\eta_n) g(m,\eta_k) - 2 h(m,\eta_n)h(m,\eta_k)) \nonumber\\
&&\hspace{-10mm}\psi^{(k)\dagger}_{x,a}\Gamma_{a,b}\psi^{(-n)}_{x,b}\psi^{(-n)\dagger}_{y,c}\Gamma_{c,d}  \psi^{(k)}_{y,d}\Big]
\;\label{case1_a1}
\eea
Here we used relations like \eq{identities} ,  \eq{eq_g_h} and
\bea
f_n f_k +f_{-n} f_{-k}&=&\nonumber\\
&&\hspace{-2cm}-2 g(m,\eta_n) g(m,\eta_k) + 2 h(m,\eta_n)h(m,\eta_k)\nonumber\\
f_{-n} f_k+f_n f_{-k}&=&\nonumber\\
&&\hspace{-2cm}-2 g(m,\eta_n) g(m,\eta_k) - 2 h(m,\eta_n)h(m,\eta_k)\;.
\eea
The propagator is
\begin{align}
P_c(\I\,\Gamma\gamma_5)&=\I^2s_{\Gamma\gamma5}\sum_{n>0,k>0} \nonumber\\
&\Big[(-2 g(m,\eta_n) g(m,\eta_k) + 2 h(m,\eta_n)h(m,\eta_k))\nonumber\\
& \psi^{(k)\dagger}_{x,a}(\Gamma\gamma_5)_{a,b}\psi^{(n)}_{x,b}\psi^{(n)\dagger}_{y,c}(\Gamma\gamma_5)_{c,d}  \psi^{(k)}_{y,d}\nonumber\\
&+(-2 g(m,\eta_n) g(m,\eta_k) - 2 h(m,\eta_n)h(m,\eta_k)) \nonumber\\
&\psi^{(k)\dagger}_{x,a}(\Gamma\gamma_5)_{a,b}\psi^{(-n)}_{x,b}\psi^{(-n)\dagger}_{y,c}(\Gamma\gamma_5)_{c,d}  \psi^{(k)}_{y,d}\Big]\nonumber\\
\end{align}
\begin{align}
\phantom{P_c(\I\,\Gamma\gamma_5)}&=-s_{\Gamma\gamma5}s_5\sum_{n>0,k>0}&& \nonumber\\
&\Big[(-2 g(m,\eta_n) g(m,\eta_k) + 2 h(m,\eta_n)h(m,\eta_k))\nonumber\\
& \psi^{(k)\dagger}_{x,a}\Gamma_{a,b}\psi^{(-n)}_{x,b}\psi^{(-n)\dagger}_{y,c} \Gamma_{c,d}  \psi^{(k)}_{y,d}\nonumber\\
&+(-2 g(m,\eta_n) g(m,\eta_k) - 2 h(m,\eta_n)h(m,\eta_k)) \nonumber\\
&\psi^{(k)\dagger}_{x,a}\Gamma_{a,b}\psi^{(n)}_{x,b}\psi^{(n)\dagger}_{y,c}\Gamma_{c,d}  \psi^{(k)}_{y,d}\Big]\;.
\end{align}
where we replaced   in \eq{propG} $s_\Gamma$ by $s_{\Gamma \gamma5}$ and $\Gamma$  by $\Gamma\gamma_5$ and utilized \eq{identities}. 

The difference between the propagators becomes in all cases
\bea
&&P_{c}(\Gamma)-P_{c}(\Gamma\gamma_5)\\
&&\qquad=-4\,\sum_{n>0,k>0} g(m,\eta_n)g(m,\eta_k)\nonumber\\
&&\qquad\qquad\qquad\Big[
\psi^{(k)\dagger}_{x,a}\Gamma_{a,b}\psi^{(n)}_{x,b}
\psi^{(n)\dagger}_{y,c}\Gamma_{c,d}  \psi^{(k)}_{y,d} \nonumber\\
&&\qquad\qquad\qquad
+\psi^{(k)\dagger}_{x,a}\Gamma_{a,b}\psi^{(-n)}_{x,b}
\psi^{(-n)\dagger}_{y,c}\Gamma_{c,d}  \psi^{(k)}_{y,d}\Big]\;.\nonumber
\eea
\section{Disconnected terms}\label{app_b}
These terms are responsible for the difference between isovector and isoscalar propagators and have  the form
\be
P_d(\Gamma)=-s_\Gamma\left[\sum_k  f_k  \psi^{(k)\dagger}_{x,a}\Gamma_{a,b}\psi^{(k)}_{x,b}\right]
\left[\sum_n f_n  \psi^{(n)\dagger}_{y,c}\Gamma_{c,d}\psi^{(n)}_{y,d}\right]\;.
\ee
Rewriting the first sum gives
\bea
&&-\sum_{k>0} [f_k  \psi^{(k)\dagger}_{x,a}\Gamma_{a,b}\psi^{(k)}_{x,b}
+f_{-k}  \psi^{(-k)\dagger}_{x,a}\Gamma_{a,b}\psi^{(-k)}_{x,b}]\nonumber\\
&&=\sum_{k>0} [f_k  \psi^{(k)\dagger}_{x,a}\Gamma_{a,b}\psi^{(k)}_{x,b}
+f_{-k} s_5 \psi^{(k)\dagger}_{x,a}\Gamma_{a,b}\psi^{(k)}_{x,b}]\nonumber\\
&&=-\sum_{k>0} (f_k +s_5 f_{-k} ) \psi^{(k)\dagger}_{x,a}\Gamma_{a,b}\psi^{(k)}_{x,b}
\eea
where we used \eq{identities} in the 2nd step. Equivalent derivation for the 2nd sum leads to
\bea
P_d(\Gamma)&=&-s_\Gamma
\sum_{k>0.n>0} (f_k +s_5 f_{-k} )(f_n +s_5 f_{-n} )\nonumber \\
&&\qquad\quad\times\psi^{(k)\dagger}_{x,a}\Gamma_{a,b}\psi^{(k)}_{x,b}
\psi^{(n)\dagger}_{y,c}\Gamma_{c,d}\psi^{(n)}_{y,d}
\eea
For $\Gamma\in\{\mathbf{1},$ $\gamma_k\gamma_j$, $\gamma_k\gamma_4$, 
$\I\,\gamma_5$, $\I\,\gamma_k\gamma_j\gamma_5$, $\I\, \gamma_k\gamma_4\gamma_5 \}$ we find $s_5=1$ and
\be
(f_k + f_{-k} )(f_n + f_{-n} ) =-4\,g(m,\eta_k)g(m,\eta_n)\;,
\ee
giving \eq{case2result}.

For $\Gamma\in\{\gamma_k,\gamma_4,\gamma_k\gamma_5, \gamma_4\gamma_5 \}$ we find $s_5=-1$ and
\be\label{disconncase3}
(f_k - f_{-k} )(f_n - f_{-n} ) =4 \,h(m,\eta_k)h(m,\eta_n)\;,
\ee
which is discussed below in App. \ref{app_c}.

\section{More disconnected terms}\label{app_c}
This concerns the disconnected terms \eq{disconncase3} in Sect. \ref{sec_more_disc_terms}.
We consider the disconnected contribution to propagators for $\Gamma\in\{\gamma_k,\gamma_4,\gamma_k\gamma_5, \gamma_4\gamma_5 \}$
discussed at the end of App. \ref{app_b}. Since the functions $h(m,\eta)$ have slower decay towards larger $\eta$ we have a closer look at the matrix elements. In the chiral basis of \eq{LRbasis} the matrices $\Gamma$ have the form
\be
\left(\begin{matrix} 0&\sigma \cr \sigma&0\end{matrix}\right)\textrm{~~or~~}\left(\begin{matrix} 0&-\sigma\cr \sigma&0\end{matrix}\right)\;.
\ee
Then in all cases we find the form ($\sigma$ depends on the actual $\Gamma$ and is proportional to a Pauli matrix)
\bea
P_d(\Gamma)&=&
4 s_\Gamma\sum_{k>0.n>0} h(m,\eta_k)h(m,\eta_n) \\
&&\quad\times
(\psiR^{(n)\dagger}(x)\,\sigma \,\psiL^{(n)}(x)+\psiL^{(n)\dagger}(x)\,\sigma \,\psiR^{(n)}(x)) \nonumber\\
&&\quad\times
(\psiR^{(k)\dagger}(y)\,\sigma \,\psiL^{(k)}(y)+\psiL^{(k)\dagger}(y)\,\sigma \,\psiR^{(k)}(y))\,,\nonumber
\eea
In all terms upper components couple to lower ones. If the overlap is small (e.g., is the eigenmodes are close to chiral) then this contribution is small and the isovector and isoscalar propagators for that $\Gamma$ are similar.
\section{$P_c(\Gamma)-P_c(\Gamma\gamma_4)$}\label{app_d}
Also the connected propagator differences between these pairs need additional assumptions like
 those of App. \ref{app_c}.
We inspect pairs $(\Gamma,\Gamma\gamma_4)$ for $\Gamma\in\{\mathbf{1}$, $\gamma_k$, 
$\gamma_5$,  $\gamma_k\gamma_j$,  $\gamma_k\gamma_5 \}$

Using \eq{case1_a1} we get
\bea\label{case3_a1}
&&\hspace{-6mm}P_{c}(\Gamma)-P_{c}(\Gamma\gamma_4)
=\sum_{n>0,k>0}\nonumber\\
&&\Big[ (-2 g(m,\eta_n) g(m,\eta_k) + 2 h(m,\eta_n)h(m,\eta_k))\nonumber\\
&&\hspace{-1mm}(s_\Gamma 
\psi^{(k)\dagger}_{x,a}\Gamma_{a,b}\psi^{(n)}_{x,b}
\psi^{(n)\dagger}_{y,c}\Gamma_{c,d}  \psi^{(k)}_{y,d}\nonumber\\
&&\hspace{-1mm}-s_{\Gamma\gamma_4} 
\psi^{(k)\dagger}_{x,a}(\Gamma\gamma_4)_{a,b}\psi^{(n)}_{x,b}
\psi^{(n)\dagger}_{y,c}(\Gamma\gamma_4)_{c,d}  \psi^{(k)}_{y,d})\nonumber\\
&&\hspace{-1mm}+(-2 g(m,\eta_n) g(m,\eta_k) - 2 h(m,\eta_n)h(m,\eta_k)) \nonumber\\
&&\hspace{-1mm}(s_\Gamma
\psi^{(k)\dagger}_{x,a}\Gamma_{a,b}\psi^{(-n)}_{x,b}
\psi^{(-n)\dagger}_{y,c}\Gamma_{c,d}  \psi^{(k)}_{y,d} \nonumber\\
&&\hspace{-1mm}-s_{\Gamma\gamma_4}
\psi^{(k)\dagger}_{x,a}(\Gamma\gamma_4)_{a,b} \psi^{(-n)}_{x,b}
\psi^{(-n)\dagger}_{y,c}(\Gamma\gamma_4)_{c,d}  \psi^{(k)}_{y,d}))
\Big]
\eea

We change to formulation
\eq{LRbasis}; the matrix pair $\Gamma$ and $\Gamma\gamma_4$ have a form like, e.g.,
\be
\left(\begin{matrix} \sigma&0\cr0& \sigma\end{matrix}\right)\textrm{~~and~~}\left(\begin{matrix} 0&\sigma\cr \sigma&0\end{matrix}\right)\;.
\ee
As example we take $\Gamma=\mathbf{1}$ and  $\Gamma\gamma_4=\gamma_4$ where $s_{\Gamma}=s_{\Gamma\gamma_4}=1$. (The other combinations give similar results, differing only in some signs.) Then \eq{case3_a1} becomes
\bea
4\sum_{n>0,k>0}\Big[ &&
g(m,\eta_k)g(m,\eta_n)\\
&&( - \psiL^{(n)\dagger}(x)\psiL^{(k)}(x) \psiL^{(k)\dagger}(y)\psiL^{(n)}(y)\nonumber\\
&& + \psiR^{(n)\dagger}(x)\psiL^{(k)}(x)  \psiL^{(k)\dagger}(y)\psiR^{(n)}(y)\nonumber\\
&&+ \psiL^{(n)\dagger}(x)\psiR^{(k)}(x) \psiR^{(k)\dagger}(y)\psiL^{(n)}(y)\nonumber\\
 &&- \psiR^{(n)\dagger}(x)\psiR^{(k)}(x) \psiR^{(k)\dagger}(y)\psiR^{(n)}(y))\nonumber\\
&&+h(m,\eta_k)h(m,\eta_n)\nonumber\\
 &&( - \psiL^{(n)\dagger}(x)\psiR^{(k)}(x) \psiL^{(k)\dagger}(y)\psiR^{(n)}(y)\nonumber\\
&&- \psiR^{(n)\dagger}(x)\psiL^{(k)}(x)  \psiR^{(k)\dagger}(y)\psiL^{(n)}(y)\nonumber\\
&&+ \psiR^{(n)\dagger}(x)\psiR^{(k)}(x)  \psiL^{(k)\dagger}(y)\psiL^{(n)}(y)\nonumber\\
&&+ \psiL^{(n)\dagger}(x)\psiL^{(k)}(x)  \psiR^{(k)\dagger}(y)\psiR^{(n)}(y))\Big]\;.\nonumber
\eea
Again we find a term with $g(m,\eta_k)g(m,\eta_n)$ which vanishes if the small modes disappear. 
The term with $h(m,\eta_k)h(m,\eta_n)$ has significant contributions from low modes which disappear 
with them. It decays, however, slower for increasing $\eta$. If the modes above the 
gap are close to chiral, this term becomes small as well. In that case we are left with the $g$-type terms
and the propagators are $g$-equivalent.


\end{document}